\documentclass[12pt]{article}

\usepackage{latexsym}
\usepackage{amsmath}
\usepackage{amssymb}
\usepackage[dvips]{graphicx}
\usepackage{latexsym}

\newcommand{\C}{\mathbb{C}}
\newcommand{\CP}{\mathbb{CP}}

\newcommand{\R}{\mathbb{R}}

\newcommand{\koniec}{\begin{flushright}  $\Box $ \end{flushright}}
\def\be{\begin{equation}}
\def\ee{\end{equation}}

\def\theequation{\thesection.\arabic{equation}}

\def\t{\tilde}

\def\sm{\sigma}

\def\Om{\Omega}

\def\p{\partial}
\def\ov{\overline}

\def\lra{\longrightarrow}

\def\ra{\rightarrow}

\def\h{\hat}

\def\om{\omega}
\def\mc{\mathcal}

\newcommand{\hook}{{\setlength{\unitlength}{11pt}   
                   \begin{picture}(.833,.8)
                   \put(.15,.08){\line(1,0){.35}}
                   \put(.5,.08){\line(0,1){.5}}
                   \end{picture}}}

\def\ll{\lambda}

\newtheorem{theo}{Theorem}[section] 
\newtheorem{prop}[theo]{Proposition}  
\newtheorem{lemma}[theo]{Lemma}

\def\theequation{\thesection.\arabic{equation}}

\begin{document}
\title{\vskip -70pt
\begin{flushright}
{\normalsize DAMTP-2010-79} \\
\end{flushright}
\vskip 10pt
{\bf Scalar--flat K\"ahler metrics with conformal Bianchi V symmetry \vskip 15pt}}
\author{Maciej Dunajski\thanks{Email: M.Dunajski@damtp.cam.ac.uk}\\
Department of Applied Mathematics and Theoretical Physics,\\
University of Cambridge,\\
Wilberforce Road, Cambridge CB3 0WA, UK. \\
{\em and}\\
Institute of Mathematics of the  Polish Accademy of Sciences,\\
\'Sniadeckich 8,
00-956 Warsaw, Poland.
\\[10pt]
Prim Plansangkate\thanks{Email: plansang@CRM.UMontreal.CA}\\
Centre de Recherches Math\'ematiques (CRM),\\
 Universit\'e de Montr\'eal, C.P. 6128,\\ 
Montr\'eal (Qu\'ebec) H3C 3J7, Canada.
\\[20pt]
{\em Dedicated to Maciej Przanowski on the occasion of his 65th birthday.}}
\date{}
\maketitle
\begin{abstract}
We provide an affirmative answer to a question posed by Tod \cite{Tod:1995b}, and construct all
four--dimensional K\"ahler metrics with vanishing scalar curvature which are invariant under
 the conformal action of Bianchi V group.
The construction is based on the combination of twistor theory and the isomonodromic problem with two double poles. The resulting metrics are 
non--diagonal in the left--invariant basis and are 
explicitly given in terms of Bessel functions and their integrals.
We also make a connection with the LeBrun ansatz, and characterise the associated solutions of the $SU(\infty)$ Toda equation by the existence a non--abelian two--dimensional group of point symmetries.
\end{abstract}
\section{Introduction}
\setcounter{equation}{0}
Let $(M, g)$ be a Riemannian four--manifold with anti--self--dual (ASD) Weyl curvature. 
The metric $g$ is said to have cohomogeneity--one 
if $(M, g)$  admits an isometry group acting transitively on 
codimension-one surfaces in $M$. We shall say that a conformal structure $[g]$ on $M$ has cohomogeneity--one, if there exists a cohomogeneity--one metric $g\in[g]$. 

The four-dimensional cohomogeneity-one metrics can be classified
according to the Bianchi type  of the three-dimensional real Lie
algebra \cite{Bianchi:1897a} of the isometry group $G$. Locally $M=\R\times G$, and 
the problem of finding ASD cohomogeneity--one metrics  reduces to solving a system of ODEs, 
as the group coordinates do not appear in the Weyl tensor. Moreover, the 
reduction--integrability dogma suggests that the resulting ODEs will be in some sense solvable as the 
underlying anti--self--duality equations are integrable by twistor transform \cite{Penrose:1976js,Mason:1991rf,Dunajski_book}.
This is indeed the case. If $G=SU(2)$ then the ODEs reduce to Painleve VI, or 
(if additional assumptions are made about the metric) 
Painleve III ODEs \cite{Tod:1994a,
Hitchin:1995a,Tod:1995a, Tod:1995b,Maszczyk:1994dt,Dancer,Dancer:1995a}.
In \cite{Tod:1995b} Tod has analysed the general case of conformal ASD 
cohomogeneity--one metrics 
with arbitrary group, diagonal in the basis of left--invariant one forms on the group. He has shown that  
cohomogeneity--one diagonal Bianchi V metrics are conformally flat. The diagonalisabilty assumption can only be made without loss of generality
 if the underlying metric is Einstein, so the existence of non--diagonal Bianchi V  conformal structures has not been ruled out in Tod's work. In particular he has raised a question whether such structures (if they exist) can admit a K\"ahler metric in a conformal class.  In this paper we shall use the method of isomonodromic
deformations to construct all such conformal structures.

Let $\mathfrak{g}$ be a Lie algebra of Bianchi type V, and let $G$ be a corresponding simply connected Lie group (see Appendix). The left--invariant vector fields on $G$ satisfy
\be 
\label{leftinvarvect} [L_0, L_1] = L_1, \quad  [L_0, L_2] = L_2,
 \quad  [L_1, L_2] = 0. 
\ee
A general cohomogeneity--one metric on $M=\R\times G$ will be have the form
\[
g=dt^2+h_{jk}(t)\ll^j\odot\ll^k, \qquad j,k=0, 1, 2
\]
where $\ll^j$ are the Maurer--Cartan one--forms on $G$ such that 
$L_j\hook \ll^k={\delta_j}^k$.

We will say that a group action on a complex manifold is holomorphic 
if Lie derivatives of the complex stucture along any of the generators vanish.
\begin{theo}
\label{complexV}
Any cohomogeneity--one Bianchi $V$ ASD conformal structure which admits a 
K\"ahler metric such that the group acts holomorphically
can be locally represented by a cohomogeneity--one metric
\be
\label{gtypeVgen}
g=e^1\odot \ov{e}^1+e^2\odot\ov{e}^2,
\ee
where the complex one--forms $e^1, e^2$ are dual to the vector fields
\[
E_1=\p_t-\frac{i}{t}(r_0L_0+r_1L_1+r_2L_2), \quad
E_2=p_0L_0+p_1L_1+p_2 L_2
\]
and the real functions ${\bf r}(t)=(r_0(t), r_1(t), r_2(t))$ and complex functions
${\bf p}(t)=(p_0(t), p_1(t), p_2(t))$ satisfy the linear system of ODEs
\be
\label{pqreqmatrix}
\frac{d{\bf r}}{dt}=t{{Im}}\;(\ov{p}_0{\bf{p}}), \quad t \frac{d {\bf p}}{dt}=i(r_0{\bf p}-p_0{\bf r}), \quad \frac{d p_0}{dt}=0, \quad \frac{d r_0}{d t}=0
\ee
such that $\mbox{det}\;[{\bf r}, {\bf p}, {\bf\ov{p}}]\neq 0$.
\end{theo}
We shall prove this theorem in Section \ref{section2}, where we shall also show that 
equations (\ref{pqreqmatrix}) reduce to the Bessel equation and quadratures.

All ASD K\"ahler metrics have vanishing scalar curvature -  we shall follow LeBrun \cite{LeBrun:1991a} in calling them scalar--flat K\"ahler  - and conversely 
all scalar--flat K\"ahler metrics have anti--self--dual Weyl tensor (we choose the natural orientation, where the K\"ahler two--form is self--dual).
In Section
\ref{section3} we shall construct the K\"ahler structure in the conformal class of Theorem
\ref{complexV}. We shall demonstrate that although a $G$--invariant conformal factor turning the metric
(\ref{gtypeVgen}) into a K\"ahler metric does not exist, there is a simple function on $G$ which does the job. We shall prove
\begin{prop}
\label{prop2}
Any conformally Bianchi--V scalar--flat K\"ahler metric can locally be put into a form
\be
\label{Kahlerconf}  
g_K = \Om^2 \; g,  
\ee
where $g$ is given by {\em(\ref{gtypeVgen})\em} and $\Omega:G\rightarrow \R$ is a function on the group such that
\be 
\label{conffacteq} 
L_0(\Omega)=\Omega, \quad L_1(\Om) = 0, \quad L_2(\Om)=0.
\ee
The holomorphic (1,0) vector fields are given by $E_1$ and $E_2$. Moreover $g_K$ is Ricci flat if and only if it is flat. 
\end{prop}
The right--invariant vector fields $R_j$ on $G$ generate the conformal transformations of the K\"ahler metric (\ref{Kahlerconf})
\[
{\cal L}_{R_0} g_K=2g_K, \quad {\cal L}_{R_1} g_K=0, \quad {\cal L}_{R_2} g_K=0,
\] 
where ${\cal L}$ denotes the Lie derivative. If we choose coordinates $(\rho, x^1, x^2)$ on $G$ (see Appendix) such that
\[
L_0=\rho\frac{\p}{\p \rho},\quad L_1=\rho\frac{\p}{\p x^1},\quad L_2=\rho\frac{\p}{\p x^2}, \qquad \rho\in\R^+, (x^1, x^2)\in\R^2
\]
then $\Omega=\rho$.

Equations (\ref{pqreqmatrix}) imply that $r_0$ and $p_0$ are constants. The remaining equations reduce to the Bessel equation possibly with imaginary order. A particularly
simple class of solutions characterised by $r_0=0$ is 
\[
{\bf r}=(0, tJ_1(t), t Y_1(t)), \quad {\bf p}=(1, iJ_0(t), iY_0(t)),
\]
where $J_\alpha(t)$ and $Y_\alpha(t)$ are Bessel  functions of first and second type of order $\alpha$. This leads to the following example of a Bianchi-V scalar flat K\"ahler metric
\begin{eqnarray} 
\label{Kahlerr00} 
g_K &=& \frac{1}{4}(d\rho^2+\rho^2 dt^2) +G_{AB}(t)dx^Adx^B,\qquad A, B=1, 2\\
&& \mbox{where} \quad
G(t)=\frac{\pi^2t^2}{16}
\left( \begin{array}{cc}
Y_0^2+Y_1^2 & -J_0Y_0 - J_1Y_1\\
-J_0Y_0 - J_1Y_1 &  J_0^2+J_1^2
\end{array} \right).\nonumber 
\end{eqnarray} 
 
 All scalar--flat K\"ahler metrics admitting a Killing vector which also preserves the K\"ahler
form  arise from  the LeBrun's ansatz \cite{LeBrun:1991a}, and 
are determined by a solution to the $SU(\infty)$-Toda field equation  
\be 
\label{Todaeq}
u_{xx}+ u_{yy}+ \left( e^u \right)_{zz} = 0, \qquad\mbox{where}\quad u=u(x, y, z),
\ee
together with a solution to its linearisation.
K\"ahler metrics arising from Proposition \ref{prop2} admit a two--dimensional group of symmetries (generated by the right translations $R_1$ and $R_2$) preserving the K\"ahler from. Thus  any linear combination of $R_1$ and $R_2$
 should lead to a solution of 
(\ref{Todaeq}). In Section \ref{section4} we shall characterise the solutions corresponding to
the  metric (\ref{Kahlerr00}). They admit a two--dimensional non--abelian
 group of Lie point symmetries and can be found by making an ansatz $u=u(y/z)$.

\vspace{2ex}{\bf Acknowledgements.} 
We are grateful to  Philip Boalch, Andrew Dancer and  
Paul Tod for helpful discussions. 
\section{ASD structures and isomonodromy}
\label{section2}
A K\"ahler structure on a four--dimensional real manifold $M$
consists of a pair $(g, I)$ where $g$ is a Riemannian
metric and $I:TM\rightarrow TM$ is a complex structure
such that, for any vector fields $X, Y$, $g(X, Y)=g(IX, IY)$
and the two--form $\omega$ defined by
\[
\omega(X, Y)=g(IX, Y)
\]
is closed. Given an orientation on $M$ the
Hodge operator $\ast:\Lambda^2\rightarrow \Lambda^2$
satisfies $\ast^2=\mbox{Id}$
and gives a decomposition
$\Lambda^{2} = \Lambda_{+}^{2} \oplus \Lambda_{-}^{2}
$
of two---forms
into self--dual (SD)
and anti--self--dual (ASD) components. The two--form $\omega$ induces a natural orientation
on $M$ given by the volume form $\omega\wedge \omega$.
With respect to this orientation 
$\omega$ is self--dual. It is well known \cite{Pontecorvo:1992a, Dunajski_book} that if the scalar curvature of a K\"ahler
metric vanishes, then the Weyl tensor of the underlying conformal structure is ASD. 
Conversely ASD K\"ahler metrics are scalar--flat.

A convenient way to
express the ASD condition on a conformal structure is summarised in
the following theorem.  The theorem below is originally due to
Penrose \cite{Penrose:1976js}, but taken in this form from \cite{Mason:1991rf,Dunajski_book}.
\begin{theo}
\label{tetradlaxprop}
Let $E_1, E_2$ be two complex vector fields in $TM\otimes\C$
and let $e^1, e^2$ be the corresponding dual one--forms. The
conformal structure defined by
\[
g=e^1\odot\ov{e}^1+e^2\odot\ov{e}^2
\]
is 
ASD  if and only if there exists functions 
$f_0, f_1$ on $M \times \CP^1$ holomorphic in $\lambda\in \CP^1$ 
such that the distribution 
\be 
\label{tetradlax0} l = \ov{E}_1 -  \ll {E}_2 + f_0 \frac{\p}{\p \ll}, \quad m =
-\ov{E}_2 -  \ll {E}_1 + f_1 \frac{\p}{\p \ll}  \ee
is Frobenius integrable, that is, $[l,m] = 0$ modulo $l$ and $m.$
\end{theo}
A general ASD conformal structure $[g]$ does not admit a K\"ahler metric in its conformal class.
The existence of such metric is characterised by vanishing of higher order
conformal invariants of $[g]$ \cite{DT10}. In what follows we shall use a simpler twistor
characterisation of ASD metrics which are conformal to K\"ahler.

\subsection{Twistors and divisors}
Let us complexify $(M, g)$ and regard $M$ as a holomorphic four--manifold with a holomorphic metric $g$
\be \label{gtetrad} 
g = \left( V^{00'} \odot V^{11'} - V^{01'} \odot V^{10'}
\right), \ee
where $V^{AA'}, A, A'=0, 1$ is the  null tetrad of one forms written in the two--component 
spinor notation. The reality conditions can be imposed to recover the Riemannian metric by setting \[
V^{00'}=\ov{e}^1, \quad V^{10'}=-\ov{e}^2,\quad   V^{01'}= {e}^2,\quad  V^{11'}={e}^1.\]
 Let $V_{AA'}$ be the corresponding tetrad of vector fields
so that the integrable distribution of Theorem \ref{tetradlaxprop} is
\be \label{tetradlax} 
l = V_{00'} -  \ll V_{01'} + f_0 \frac{\p}{\p \ll}, \quad m =
V_{10'} -  \ll V_{11'} + f_1 \frac{\p}{\p \ll}.  
\ee
A twistor space of $(M, g)$ is the space of two--dimensional totally null surfaces spanned by 
$V_{00'} -  \ll V_{01'}, V_{10'} -  \ll V_{11'}$ in $M$. It is a three dimensional complex manifold ${\cal Z}$ which arises as a quotient of $M\times\CP^1$ by ${l, m}$.
The points in $M$ correspond to rational curves (called twistor curves) in ${\cal Z}$ with normal  bundle
${\cal O}(1)\oplus{\cal O}(1)$, where ${\cal O}(n)$ is a line bundle over $\CP^1$ with Chern class $n$. 
The holomorphic canonical line bundle
$\kappa$ of ${\cal Z}$ restricted to any of these curves is isomorphic to ${\cal{O}}(-4)$.
To reconstruct a real four--manifold $M$, the twistor curves must be invariant under an anti--holomorphic involution $\tau$ on ${\cal Z}$ which restricts to an antipodal map on each curve.

A theorem of Pontecorvo \cite{Pontecorvo:1992a} 
states that an ASD conformal structure $(M, [g])$ admits a  K\"ahler metric if and
only if there exists a section $D$ of the bundle ${\mc O}(2) = \kappa^{-1/2}$ over the
twistor space ${\cal Z}$ of $(M, [g]),$ with exactly two distinct zeros
on each  twistor line. The section must be $\tau$ invariant, so its zeroes lie on the antipodal points on each twistor curve.

Now consider the group $G$ acting on an ASD conformally K\"ahler manifold $M$ by holomorphic conformal isometries with generically three--dimensional orbits. This gives rise to a complexified group action of $G_\C$ on 
the twistor space ${\cal Z}$. Let $\widetilde{R}_j, j=0, 1, 2$ be holomorphic vector fields
on ${\cal Z}$ corresponding to the right--invariant conformal Killing vectors $R_j$ on $M$.
The subset of ${\cal Z}$ where $\widetilde{R}_j$ are linearly dependent is given by
the zero set of $s=vol_{\cal{Z}}(\widetilde{R}_0, \widetilde{R}_1, \widetilde{R}_2)$.
As the canonical bundle $\kappa={\cal{O}}(-4)$, the divisor $s=0$ defines a quartic and vanishes at four points on each twistor line. Hitchin (\cite{Hitchin:1995a}, Proposition 3)
showed\footnote{His result was derived
for $G_\C=SL(2, \C)$ but it remains valid for any three-dimensional group acting transitively on $M$, as the corresponding action of $G_\C$ preserves the points where Pontecorvo's divisor vanishes as long as the group action is holomorphic. Moreover we are allowed to work
in a conformal class of $g_K$, as the twistor equation underlying the Pontecorvo's divisor
is conformally invariant \cite{Pontecorvo:1992a, DT10}.} 
that $s$ is not identically zero if $g$ is not Ricci--flat and that the divisor 
when $s$ vanishes is equal to the Pontecorvo's divisor $D$ in the case when $[g]$ is cohomogeneity one and contains a K\"ahler class. In the double fibration picture
\[
M\longleftarrow M\times\CP^1\longrightarrow {\cal Z}
\]
the section $s$ pulls back to 
\be 
\label{cxKo2section} s = (d \ll \wedge \nu) (l, m,  \widetilde{R}_0,  \widetilde{R}_1,
 \widetilde{R}_2), \ee
where $ \nu =  V^{01'} \wedge V^{10'}  \wedge V^{11'} \wedge V^{00'}$ is the volume form 
on $M$
and
 $\widetilde{R}_j$ are the lifts of the three generators of
$G$ to $M \times \CP^1$ such that $[l, \widetilde{R}_j] = 0,$ $[m, \widetilde{R}_j] = 0$ modulo $l, m,$. Thus in the proof of Theorem \ref{complexV} we will require that this quartic has  
two distinct zeros of order two. This will guarantee the existence of a K\"ahler metric 
in the cohomogeneity--one conformal class.

\vspace{0.5cm}

\noindent {\bf Proof of Theorem \ref{complexV}.}
We will work in the complexified category and impose the reality conditions at the end.
Let $G_\C$ be a three-complex dimensional Lie group
(we will eventually take $G_\C$ to be a complexification of the Bianchi V group, but the first part of the proof does not depend on the choice of $G_\C$).
We assume that the orbits of $G_\C$ are three--dimensional, and the 
metric $g$  on $\C \times G_\C$  is invariant
under the left translations of $G_\C$ on itself.  One can write the null tetrad 
$V_{AA'}$ in
terms of the vector field $\p_t$ and  three linearly
independent vector fields $P, Q, R$ tangent to $G_\C$ 
(in the complexified setting the fields $Q$ and $P$ are independent. Once the reality conditions are imposed at the end of the proof, we shall set $Q=-\ov{P}$)
which are
$t$-dependent and invariant under the left translations, as
\be \label{tetrad}  V_{00'} = \p_t +i \frac{R}{t}, \quad V_{11'} = \p_t -i \frac{R}{t}, 
\quad V_{01'} = P,
\quad V_{10'}=Q.  \ee  

Now, let $R_0, R_1, R_2$ be the right-invariant vector fields on
$G$ corresponding to three generators of the left translations. 
Since $R_j$ are independent of $t,$ one has
\[ [l,  R_j] = - R_j(f_0) \p_\ll, \quad   [m,  R_j] = -  R_j(f_1)
\p_\ll. \]
A direct calculation shows that there is no lift of $ R_j$ of the form
$ R_j + {\cal Q}_j \p_\ll$ for some function ${\cal Q}_j$ such that 
$ [l, R_j + {\cal Q}_j \p_\ll] = 0, \;  [m, R_j + {\cal Q}_j \p_\ll] = 0$  modulo $l,
m$. Hence, we conclude that $[l, 
  R_j], [m,  R_j],$ are identically zero.  
This implies
that $f_0$ and $f_1$ are constant on $G,$ and hence they are functions
of $\ll$ and  $t$ only.

We claim that the  quartic $s$ 
as defined in (\ref{cxKo2section}) is proportional to $\ll
f_0 + f_1,$ with the proportionality factor given by a function $h$ on $M.$ Indeed, using the fact that
$V^{01'},\; V^{00'} - V^{11'},\; V^{10'}$
are linearly independent and invariant under the left translations of $G_\C,$  the 
right-invariant vector fields $R_j $ can be written in the 
basis of $ P = V_{01'}, \; Q = V_{10'}, \; R = \frac{it}{2} (V_{11'} - V_{00'}).$ 
Thus, performing all the contractions we are left with
$ s  \propto (\det H) \; (\ll
f_0 + f_1)$, 
where $H$ is the matrix of coefficients of $ R_0,  R_1,  R_2$ written
in the basis of $P, Q, R.$  

\vspace{0.5cm}

As $\det H$ does not depend on
$\ll,$  the quartic $s$ has two distinct zeros of order two if and only if 
$\ll f_0 + f_1$  has two distinct zeros of order two (for the moment we rule out the case where 
$f_0$ and $f_1$ are both zero which corresponds to a hyper-K\"ahler metric \cite{Dunajski_book}).  
We shall assume that this is the case so that the ASD conformal structure
admits a K\"ahler metric. It is now possible to use M\"obius
transformation to put the two zeros at $0$ and $\infty.$  The M\"obius
transformation in $\ll$ corresponds to a change of null tetrad by
a right rotation $V_{AA'} \ra V_{AA'} {r^{A'}}_{B'},$ where $r$ is an
$SL(2, \C)$-valued function.  Since the coefficients of the quartic
$\ll f_0 + f_1$ are functions of $t$ only, the required ${r^{A'}}_{B'}$ will only
depend on $t.$  Thus, the rotated tetrad is still $G_\C$-invariant.

With the zeros at $0$ and $\infty,$ $\ll f_0 + f_1$ is of the form
$a(t) \ll^2.$   Let us first assume that $a$ does not vanish identically. One 
 still has a M\"obius degree of freedom that preserves $(0,
\infty),$ that is, the multiplication of $\ll$ by
a function of $t.$   Let us use this freedom to set $a(t) = 2/t.$
The current tetrad is some right rotation of the original one.  It is
possible to use another freedom: a left rotation $V_{AA'}
\ra {l^A}_B V_{AA'}, \; l \in SL(2, \C)$ to keep $V_{00'} - V_{11'}, V_{01'}, V_{10'}$
tangent to $G_\C.$  The right rotation does
not change the quartic $\ll f_0 + f_1,$ and we now have $f_0, f_1$ of
the form
\be \label{f0f1}
f_0 = b(t) \ll^2 + c(t) \ll + d(t), \quad
f_1 = -b(t) \ll^3 +\left(\frac{2}{t}-c(t)\right) \ll^2 - d(t) \ll
\ee
for some functions $b(t), c(t), d(t).$  

Now, consider a pair of linear combinations of $l$ and $m$ (\ref{tetradlax})
\begin{eqnarray}
\label{tetradisolaxL}
L &=& \frac{\ll l + m}{\ll f_0 + f_1}\; = \; \frac{\p}{\p \ll} + \frac{2
  \ll iR t^{-1} - \ll^2 P +Q}{\ll f_0+ f_1}  \\
M &=& \frac{f_1 l - f_0 m}{\ll f_0 + f_1} \; = \; \frac{\p}{\p t} +
  \frac{(f_1 - \ll f_0) iR t^{-1} - \ll f_1 P - f_0 Q}{\ll f_0 +f_1}.\nonumber
\end{eqnarray}
Since the conformal class is ASD, Theorem \ref{tetradlaxprop} means  
$[l, m] = (\dots)l + (\dots)m.$  This in turn implies that $[L, M] =
0,$ modulo $L$ and $M.$  However, one sees that $[L,M]$ does not
contain $\p_\ll$ or $\p_t,$ thus $[L,M]$ must be identically zero.
It turns out that  $[L,M] = 0$ implies that $b(t) = 0 = d(t).$  

The compatibility conditions $[L,M] = 0$ are then given by 
\begin{eqnarray}
t P_t - i[R, P] + (t c(t)-1)P &=& 0, \nonumber \\
\label{cPQR}
2iR_t - t [P,Q] &=& 0, \\
t Q_t + i[R, Q] - (t c(t) -1)Q &=& 0. \nonumber
\end{eqnarray}
This shows that a cohomogeneity-one metric (\ref{gtypeVgen}), in 
the tetrad (\ref{tetrad})
is ASD if the vector fields $P, Q, R$  satisfy the system (\ref{cPQR}), where $c(t)$ is 
defined in (\ref{f0f1}).  Now, let 
\be \label{candwithout} \h R = R, \quad  \h P = h(t) P, \quad \h Q = h^{-1}(t) Q, \quad
\mbox{where} \quad h(t) = e^{\int \left( c(t) - \frac{1}{t} \right) dt}. \ee
The system (\ref{cPQR}) implies that the 
vector fields $\h P, \h Q, \h R$ satisfy
\begin{eqnarray}
t P_t - i[R, P] &=& 0, \nonumber \\
\label{PQR}
2iR_t - t [P,Q] &=& 0, \\
t Q_t + i[R, Q] &=& 0, \nonumber
\end{eqnarray}
where we have dropped the hat from the rescaled vector fields.   Moreover, the
tetrad (\ref{tetrad}) constructed from a solution $(P, Q, R)$ of (\ref{PQR})
gives the same metric (\ref{gtypeVgen}) as the tetrad determined from $(h^{-1}(t)P,\;
h(t)Q, \; R)$ which satisfy
(\ref{cPQR}) with $c(t) = \frac{h_t}{h} + \frac{1}{t}.$  Neither the metric nor the equations (\ref{PQR}) depend on $h$, so we can set it equal to $1$. The resulting Lax pair
(\ref{tetradisolaxL}) is
\be
\label{PIIIlax} L = \frac{\p}{\p \ll} + \frac{(t Q + 2i\ll R - \ll^2 t P)}{2\ll^2}, \qquad
 M = \frac{\p}{\p t} -  \frac{(\ll Q + \ll^3 P)}{2\ll^2},
\ee
where we again dropped  the hat from the rescaled vector fields. 
\vskip5pt
We conclude that any cohomogeneity-one metric  
(\ref{gtypeVgen}), which belongs to an ASD conformal structure admitting
a quartic $s$  defined in (\ref{cxKo2section}) with two distinct
zeros of order two, can be written in terms of a null tetrad (\ref{tetrad}),
where the vector fields $P, Q, R$ satisfy the system (\ref{PQR}). 

The three vector fields $P, Q, R$  can be written in the basis of left-invariant vector fields
$L_0, L_1, L_2$ satisfying (\ref{leftinvarvect}) as
\begin{eqnarray}
P &=& p_0(t) \;L_0 + p_1(t) \;L_1 +  p_2(t) \;L_2 \nonumber \\
\label{vectPQR}
Q &=& q_0(t) \;L_0 + q_1(t) \;L_1 + q_2(t) \;L_2, \\
R &=& r_0(t) \;L_0 + r_1(t) \;L_1 + r_2(t) \;L_2, \nonumber 
\end{eqnarray}
for some functions $p_j(t), q_j(t), r_j(t)$.
Then using the commutation relation (\ref{leftinvarvect}) the 
 system (\ref{PQR}) implies that $p_0, q_0, r_0$ are constant and  that
\be \label{pqr1} t(p_j)_t = i(r_0p_j - p_0r_j), \quad t(q_j)_t =i(-r_0q_j +q_0 r_j), 
\quad 2i(r_j)_t = t(p_0q_j-q_0p_j). 
\ee
The reality conditions corresponding to Riemannian metrics come down to choosing 
$R$ real (so that $(r_0, r_1, r_2)$ are all real functions) and $Q=-\ov{P}$
so that $q_j=-\ov{p}_j)$.
Equations (\ref{pqr1}) then give the linear system (\ref{pqreqmatrix}). 

 Let us now return to the case $a=0$ which corresponds to $f_0$ and $f_1$ vanishing in the distribution (\ref{tetradlax}).
The resulting conformal structure must then be hyper-Hermitian 
\cite{Dunajski_book}
and (as the vector fields in  
(\ref{tetradlax}) are volume preserving) it is actually conformal to 
hyper-K\"ahler.
The integrability 
$[l,m] = 0$ modulo $l,m$ implies the system of Nahm equations
\be
\label{PQRhyperK}
 P_t - i[R/t, P] =0,\quad
2i(R/t)_t - [P,Q] = 0,\quad
 Q_t + i[R/t, Q] = 0. 
\ee
Imposing the reality condition that $R$ is real and $Q=-\ov P,$
(\ref{PQRhyperK}) become
\[
\frac{d({\bf r}t^{-1})}{dt}={{Im}}\;(\ov{p}_0{\bf{p}}), \quad t \frac{d {\bf p}}{dt}=i(r_0{\bf p}-p_0{\bf r}), 
\quad \frac{d p_0}{dt}=0, \quad \frac{d (r_0t^{-1})}{d t}=0.
\]  
The  general solutions for ${\bf r}$ and ${\bf p}$ 
are given in terms of
trigonometric and exponential functions depending on the constants $p_0$ and $r_0/t$, and the resulting metric is conformally flat
(in the proof of Proposition \ref{prop2} we shall find the conformal factor which makes it flat).
\koniec

Our derivation of (\ref{PQR}) and (\ref{PIIIlax}) did not depend on the choice of the isometry group. The system (\ref{PQR})
describes the general isomonodromic deformation equations with two
double poles.
The corresponding Lax pair 
(\ref{PIIIlax}) with  $P, Q, R$ given by $2 \times 2$ matrices was shown by
Jimbo and Miwa \cite{Jimbo:1981a} to give rise to the Painlev\'e III
equation. The same Lax pair  also arises as the reduced Lax pair of the ASDYM equation, by
the Painlev\'e III group.  It is shown to be the isomonodromic Lax
pair for the Painlev\'e III equation when the
gauge group of the ASDYM connection is $SL(2, \C)$
\cite{Mason:1993a}, or (with certain algebraic constraints on normal forms)  $SL(3, \C)$ \cite{DP08}.

\subsection{Bessel equation}
For generic values of the constants $(r_0, p_0),$  the solutions to
(\ref{pqreqmatrix}) are determined by two linearly independent solutions of the 
Bessel equation\footnote{The relation with the Bessel equation is already
expected from the result of  \cite{Maszczyk:1996a}, who classified all reductions of anti--self--dual Yang Mills equations leading (by switch map) to
cohomogeneity--one  ASD conformal structures without however determining a K\"ahler class.}.

If $p_0=0$ then ${\bf r}$ is a constant vector and equations for $p_1, p_2$ can be easily integrated. The resulting metric is conformally flat. 

If $p_0\neq 0$ we rescale the metric by a constant $|p_0|^2$ and redefine $({\bf r}, p_1, p_2, t)$ to set $p_0=1$. Differentiating the first set of equations in (\ref{pqreqmatrix}) twice, and using  the second set of equations shows that the real functions  $f_1 = 2t^{-1}{(r_1)}_t$ and $f_2 = 2t^{-1}{(r_2)}_t$ satisfy a pair of Bessel equations 
\be 
\label{realBessel} 
t^2\frac{d^2 f_k}{dt^2}+t\frac{d f_k}{d t}+(t^2+r_0^2)f_k=0, \quad k=1,2.
\ee

 If $r_0\neq 0$  the general solution of
the Bessel equation (\ref{realBessel}) is given in terms of Bessel
functions of pure imaginary order $ir_0$
\[ 
f_1 = c_1  J_{i r_0}(t) + c_2  Y_{i r_0}(t), \qquad
f_2 = c_3  J_{i r_0}(t) + c_4  Y_{i r_0}(t).
\]
The constant complex coefficients $c_1, c_2, c_3, c_4$ can be
chosen so that the functions $f_1$ and $f_2$ are real, see for example \cite{Dunster:1990a}. 
Given functionally independent $f_1$ and $f_2$  we find $r_1, r_2, p_1, p_2$ 
by integrations and algebraic manipulations.

The case $r_0=0$ is special. The linear system (\ref{pqreqmatrix})
now reduces to a pair of ODEs
\[
t^2\frac{d^2 r_k}{d t^2}-t\frac{d r_k}{dt}+t^2r_k=0, \quad k=1, 2
\]
whose solutions are given by 
\[
r_1=c_1tJ_1(t)+c_2t Y_1(t), \quad r_2=c_3tJ_1(t)+c_4t Y_1(t),
\]
where $J_\alpha$ are $Y_\alpha$ are Bessel functions of the first and second kind respectively of order $\alpha$, and $c_1, \dots, c_4$ are real constants of integrations. In this case $p_1, p_2$ must be purely imaginary and the recursion relations
\be
\label{Bessel_id}
\p_t J_0=-J_1, \quad \p_t Y_0=-Y_1, \quad \p_t(tJ_1)=tJ_0, \quad
 \p_t(tY_1)=tY_0
\ee
imply that 
\be 
\label{ppconst} p_1 = i (c_1 J_0(t) + c_2 Y_0(t)), \quad  p_2 = i(c_3 J_0(t) + c_4 Y_0(t)).
\ee
Performing a linear transformation of the vector fields $L_1$ and $L_2$
we can always set $c_1=1, c_2=0, c_3=0, c_4=1$ which yields 
\[
{\bf r}=(0, tJ_1(t), t Y_1(t)), \quad {\bf p}=(1, iJ_0(t), iY_0(t)).
\]
To write down the metric we invert the vector fields $E_1, E_2$ from Theorem \ref{complexV} and use the fact that $2t(Y_0J_1-J_0Y_1)={4}/{\pi}$.
This yields
\begin{eqnarray*} 
g &=& \frac{1}{4}(dt^2+(\ll^0)^2) +G_{AB}(t) \ll^A \ll^B,\quad\mbox{where}\quad A, B=1,2\nonumber \\
G(t)&=&\frac{\pi^2t^2}{16}
\left( \begin{array}{cc}
Y_0^2+Y_1^2 & -J_0Y_0 - J_1Y_1\\
-J_0Y_0 - J_1Y_1 &  J_0^2+J_1^2
\end{array} \right),
\end{eqnarray*} 
where $\ll^0, \ll^1, \ll^2$ are the left--invariant one forms on $G$ which satisfy (\ref{appendixa}).
\section{K\"ahler structure}
\label{section3}
{\bf Proof of Proposition \ref{prop2}}.
The K\"ahler structure on $M$ can be read off from the divisor (\ref{cxKo2section}). In the proof of Theorem
(\ref{complexV}) we have moved the double zeros of $s$ to $0$ and $\infty$, which in spinor notation
means that $\omega_{A'B'}=o_{(A'}\iota_{B')}$ and $s\approx (\omega_{A'B'}\pi^{A'}\pi^{B'})^2$
where $\pi^{A'}$ are the homogeneous coordinates on $\CP^1$ such that $\lambda
=-\pi^{0'}/\pi^{1'}$, and the spinor basis is $o_{A'}=(0, 1), \iota_{A'}=(-1, 0)$. 
Thus, in the null tetrad (\ref{tetrad}), the K\"ahler form is proportional to
\[
\hat{\omega}=\frac{i}{2}\varepsilon_{AB}\omega_{A'B'} V^{AA'}\wedge V^{BB'}=
\frac{i}{2} (e^1 \wedge \ov e^1 + e^2 \wedge \ov e^2),
\]
and the space $T^{1, 0}(M)$ of holomorphic vector fields  on $M$ is spanned by
$V_{11'}, V_{01'}$ (equivalently by the vectors $E_1$ and $E_2$ in Theorem \ref{complexV}).
The Frobenius integrability conditions 
\[
[T^{1, 0}, T^{1, 0}]\subset T^{1, 0}
\]
guaranteeing the vanishing of Nijenhuis torsion follows from the construction,
but we can also verify it directly as
\[
[E_1, E_2]=\Big(\frac{d{p}_1}{dt}-\frac{i}{t}r_0p_1+\frac{i}{t}r_1p_0\Big)L_1
+\Big(\frac{d{p}_2}{dt}-\frac{i}{t}r_0p_2+\frac{i}{t}r_2p_0\Big)L_2=0,
\]
where we have used the constancy of $(r_0, p_0)$ and equations (\ref{pqreqmatrix}).

To determine the conformal factor we look for a function $\Omega:M\rightarrow \R$ such that
$
d(\Omega^2\hat{\omega})=0
$.
Once this has been found the K\"ahler metric $g_K$ and the associated two--form $\omega$ 
will be given by
\[
g_K=\Omega^2 g, \quad \omega=\frac{i\Omega^2}{2}
(e^1 \wedge \ov e^1 + e^2 \wedge \ov e^2).
\]
It can be verified by explicit calculation  that there is no $G$--invariant $\Omega$ (i.e. there is no conformal factor which depends only on $t$). To 
demonstrate that $\Omega:G\rightarrow \R$
such that  (\ref{conffacteq}) holds gives the correct conformal factor, consider 
\be
\label{closeomega}
2\Omega d\Omega\wedge\hat{\omega}+\Omega^2 d\hat{\omega}=0.
\ee
Since $\p_t \Om =0,$ one can write $d \Om$ in the basis of left-invariant one-forms
\[ d \Om = L_0(\Om) \ll^0 +  L_1(\Om) \ll^1 + L_2(\Om) \ll^2. 
\]
The three-form $d \hat{\om}$ can be calculated using the expressions  for the 
 dual one-forms of  
$E_1$ and $E_2$ in Theorem \ref{complexV}
and the 
Maurer-Cartan's structure equation (\ref{appendixa}).  Finally, the system
(\ref{pqreqmatrix}) is used to simplify the LHS of (\ref{closeomega}), and one finds that
(\ref{closeomega}) is satisfied if and only if (\ref{conffacteq}) holds. This conformal factor is in fact unique - it could have also been 
read off from the divisor as it is proportional to a power of $\omega_{A'B'}\omega^{A'B'}$ \cite{DT10}.

In the conformally flat 
case where $s$ in (\ref{cxKo2section}) is identically zero, 
which corresponds to the system (\ref{PQRhyperK}),  
the same conformal factor makes $g$ flat. Moreover we verify by explicit calculation that $g_K$ given by (\ref{Kahlerconf}) where $s\neq 0$
is Ricci flat if and only if it is flat, which proves the last part the Proposition.
\koniec
The conformal Killing vectors generating the group action on $g_K$ are given by
the right--invariant vector fields on $G$. If the coordinates are chosen for the group, these vectors are given by (\ref{right_inv}). \\
\noindent {\bf Example.} The simplest explicit example of the scalar--flat K\"ahler metric corresponds to $r_0=0$ in Theorem \ref{complexV} and is given by (\ref{Kahlerr00}).
The determinant of the metric (\ref{Kahlerr00}) given by
\[ \det g_K =  \frac{\pi^2\rho^2t^2}{1024}. \]
Since by definition $\rho \ne 0,$ $g_K$ may be degenerate only at $t=0$
or $t=\infty.$   The Ricci scalar $R$ is 
identically zero because the K\"ahler metric is ASD. The remaining 
curvature invariants are
\[ 
R_{abcd}R^{abcd} = \frac{256}{\rho^4t^2}, 
\qquad W_{abcd}W^{abcd}=\frac{128}{\rho^4 t^2}
\]
which indicates that $t=0$ is a singularity. Rescaling the metric by a 
conformal factor $\rho^{-2}(tf(t))^{-1}$ gives  
$W_{abcd}W^{abcd}=128f^2$, which needs to be regular if the conformal class
contains  a complete metric, but the regularity of the conformal factor
requires that $tf(t)$ is also regular and non-zero. Thus the norm of the Weyl
tensor blows up at $0$.
The asymptotic behaviour of $g_K$
for large $t$ is
\[
g_K=\frac{1}{4}\Big(d\rho^2+\rho^2 dt^2\Big)+\frac{\pi t}{8}(d(x^1)^2+d(x^2)^2).\]

\section{$SU(\infty)$ Toda equation}
\label{section4}
LeBrun \cite{LeBrun:1991a} has shown that any K\"ahler
metric $g_K$ with symmetry preserving the K\"ahler form admits a local coordinate system $\{\tau, x, y, z\}$  such that
\be 
\label{LeBrunansatz} g_K =W h +
\frac{1}{W}(d \tau + \theta)^2,  \quad \om = W e^u d x \wedge d  y + d  z \wedge (d\tau  + \theta), \ee
where 
\be
h=e^u(dx^2+dy^2)+dz^2.
\label{EWmetric}
\ee
Here $\tau$ is a coordinate along the orbits of the Killing
vector $K = \p_{\tau},\; $  $\{x, y,  z\}$ are coordinates on the
space of orbits, and $(u, W$) and $\theta$ are  functions and a one--form on
the space of orbits such that $u$ satisfies the 
$SU(\infty)$ Toda equation (\ref{Todaeq}),
$W$ satisfies the so--called monopole equation
\be \label{monopoleeq} W_{x x}+ W_{y y}+ (We^u)_{ z z} = 0  \ee
and $\theta$ is determined by $W$ together with the condition $d\omega=0$.

The ansatz 
(\ref{LeBrunansatz}) can be understood as follows.  Given that $K=
\p_{\tau}$ is a Killing vector, the metric necessarily takes the form 
(\ref{LeBrunansatz}),  
where $\frac{1}{W} =  g_K(K, K)$ and $h$ is a metric on the three--dimensional space
of orbits.  Now, since the K\"ahler form $\omega$  Lie derives  along $K,$ we have
\be \label{ztilde} 
K\hook\omega=dz
\ee 
for some function $z$ on the
space of orbits of $K$.  The isothermal
coordinates $x, y$ on the orthogonal complement of the space
spanned by $K$ and $I(K)$ (where $I$ is the complex structure) can be used together with $z$
to parametrise the space of orbits.  The metric then takes the form (\ref{LeBrunansatz})
with $h$ given by (\ref{EWmetric}) for some $u=u(x, y, z)$.
The integrability of the complex structure and the closure of the
$\omega$ imply (\ref{monopoleeq}).
The scalar-flat condition gives (\ref{Todaeq}).
\subsection{Bessel solutions to $SU(\infty)$ Toda equation}
The scalar--flat K\"ahler metric (\ref{Kahlerr00}) has two Killing
symmetries $\p/\p x^1$ and $\p/\p x^2$  preserving the K\"ahler form. We can follow the algorithm described above and  find the solution $u$ of the $SU(\infty)$ Toda
equation and the associated monopole $W$ corresponding to a linear combination
\[
K=c_1\p/\p x^1+c_2\p/\p x^2.
\]
We shall set $c_2=0$ for simplicity. Set $\tau= x^1$  so that the vector field $K = \p_{\tau}$ and 
$dx^1 = d\tau$.  Then the metric (\ref{Kahlerr00}) takes 
the form (\ref{LeBrunansatz}),
where
\begin{eqnarray}
\label{monopole_W}
h &=&\frac{1}{4W}\Big(\rho^2 dt^2 + d\rho^2 \Big)+ \frac{t^2\pi^2}{64} \; d(x^2)^2,\nonumber \\
W &=& \frac{16}{t^2\pi^2 (Y_0^2 + Y_1^2)}, \quad \mbox{and} \quad \theta = 
- \frac{(J_0 Y_0 + J_1 Y_1)}{Y_0^2 + Y_1^2} \; d(x^2).
\end{eqnarray}
Using identities (\ref{Bessel_id})  and
contracting the K\"ahler form of  (\ref{Kahlerr00}) with 
$\p/\p x^1,$ one recovers the equations (\ref{ztilde}) for $z$. The other two
coordinates are\footnote{For a Killing vector given by a
general linear combination of $\p/\p x^1$ and $\p/\p x^2$, a linear combination of Bessel functions appears in the final formula, 
and in particular
\[ 
\frac{z}{y} = -t \left(\frac{c_2 J_1 - c_1 Y_1}{c_2 J_0 - c_1 Y_0}\right). 
\]
}
\be
\label{reducetoLeBrun}
\tau  = x^1, \quad  x = -\frac{\pi x^2}{8}, \quad
 y = -\frac{\pi\rho Y_0}{8}, \quad  z =  \frac{\pi\rho tY_1}{8}.
\ee
The solution to $SU(\infty)$ Toda equation is now implicitly given by
\be
\label{Toda_prim}
e^u=t^2.
\ee
Formulae (\ref{reducetoLeBrun}) imply that $u=u(v)$, where
$v=z/y$. Thus $u$ is constant on the plane $y\;v(u)-z=0$ (compare 
\cite{Tod:1995a} where solutions constant on quadrics were constructed)
and is invariant under a two--dimensional group of 
Lie point symmetries generated by vector
fields 
\be
\label{TodaLiesym}
\p/\p x, \qquad x\p_x+y\p_y+z\p_z.
\ee
We shall now show that the existence of these symmetries uniquely 
characterises 
(\ref{Toda_prim}). Any solution $u$ of (\ref{Todaeq}) 
which is invariant under symmetries generated by 
the vector fields (\ref{TodaLiesym}) is a function $u=u(v)$. Then (\ref{Todaeq}) becomes an ODE 
\[ 
(v^2 + e^u)u_{vv} +2v u_v + u_v^2 e^u =0. \]
Equivalently, interchanging the dependent and independent variables we have
\[
(v^2 + e^u)^2 \p_u \left( \frac{v_u}{v^2 +e^u} \right) = 0,
\]
which integrates to
\be
\label{1stODE}
v_u = c (v^2 +e^u)
\ee
for some constant $c$. Now we shall argue that this constant $c$ can always be set to $-1/2,$ provided that we
consider solutions to the  $SU(\infty)$ Toda  equation as equivalent if they determine
the conformally equivalent metrics (\ref{EWmetric}).
First,  note that a transformation
\be 
\label{EWsym} u \lra u + 2 \beta, \qquad  z \lra \pm e^\beta  z,
\ee
where $\beta$ is a constant is  
a symmetry of (\ref{Todaeq}) which rescalles the three--metric by
a constant factor. Now, under (\ref{EWsym}), the variable $v$ transforms as 
$v \lra \pm e^\beta v.$  This can be used to set $c = -1/2.$

The equation (\ref{1stODE}) with $c=-1/2$ is equivalent to the Bessel equation of order 0 
\be \label{0Bessel} t^2 Y_{tt} + t Y_t + t^2 Y = 0.  \ee
To see it use 
$t^2 = e^u $ and $v = 2 \p_u \ln Y$.
 The four--dimensional metric (\ref{LeBrunansatz}) resulting from the solution $u=u(z/y)$
together with the monopole (\ref{monopole_W}) admits a conformal action of the Bianchi V group
generated by
\be \label{Todarightinv} R_0 =  \tau  \p_{\tau} + x \p_{ x}+ y \p_{ y}+ z \p_{ z}, 
\qquad R_1 = \p_{\tau },
\qquad R_2 = \p_{x }.
\ee
We shall now establish that given a solution $u=u(z/y)$ to the $SU(\infty)$ Toda equation, 
the function $W$ given by (\ref{monopole_W}) is  (up to gauge transformation) the only solution to the linearised $SU(\infty)$ Toda equation (\ref{monopoleeq})
such that the resulting metric (\ref{LeBrunansatz}) in four dimensions defines a cohomogeneity--one
Bianchi V conformal class. To show it, observe that the metric $h$ in (\ref{EWmetric})
is invariant under $R_1, R_2$ and conformally invariant $h\rightarrow c^2 h$ under
the one parameter group of transformations generated by $R_0$.
This implies that 
the function $W$ must be invariant under the Bianchi V group $G$ and the one-form $\theta$
is invariant under the translations  generated by  $R_1, R_2,$  but transforms as 
$\theta \ra c \; \theta$ under the action generated by  $R_0.$  Thus the metric (\ref{LeBrunansatz}) is conformally
invariant under $G$ if and only if the function $W$ and the components 
of the one-form $\theta$ are functions of $t$ only.
Therefore the PDE (\ref{monopoleeq}) for $W$ becomes
a second order ODE 
\[ 
W_{tt}v_t(v^2+t^2) + 2W_t(v_t(2t+vv_t)-v_{tt}(v^2+t^2))+2W(v_t-t v_{tt}) = 0. 
\]
The invariance properties of $\theta$ lead to one further constraint
\be \label{constraintW} W_t (t^2+v^2) +2 t W = 0. 
\ee
The relation (\ref{reducetoLeBrun})  yields $v = - t \frac{Y_1}{Y_0}$ and 
using  (\ref{Bessel_id}) we find
the  general solution $W$ of (\ref{constraintW}) 
\be 
\label{kW} W = \frac{k}{t^2(Y_0^2+Y_1^2)}, \quad k=\mbox{const}.
\ee
Changing the proportionality constant $k$ in $W$
amounts to changing the coordinate $\tau \lra \tau / k$ in (\ref{LeBrunansatz}) and 
rescaling the metric by the constant $k.$  Hence we can always set $k = 64/\pi^2,$ 
which gives the monopole $W$  in (\ref{reducetoLeBrun}).

{\bf Example: ASD Einstein metric with symmetry.}
Here we shall present an example of a  metric (\ref{LeBrunansatz}) 
which is obtained from a monopole $W$ different from (\ref{kW}).  

It was shown in 
\cite{Tod:1997} that any ASD Einstein metric with symmetry and non-zero scalar curvature 
can be written as
\be \label{Einsteinmet} g_E = \frac{W}{ z^2} \left[ e^u (d x^2 + d y^2) + d  z^2 \right] +
\frac{1}{W  z^2}(d \tau + \theta)^2,\ee
where
$W = \mbox{const} \; ( z u_{ z} -2)$,
and $u$ is a solution to the $SU(\infty)$ Toda  equation.
Take $u$ given by (\ref{Toda_prim}) and the Einstein  monopole $W$ with const$=1/2$. Then
\[ W = \frac{Y_0Y_1}{t(Y_0^2+Y_1^2)} -1, \quad \theta = \left[ \frac{1}{2} \left(\frac{Y_0^2-Y_1^2}{Y_0^2+Y_1^2} \right) + \ln \left( \frac{\pi \rho}{8} \right) 
\right] \; d  x. \]
The resulting Einstein metric (\ref{Einsteinmet}) has negative scalar curvature and is non-conformally flat.   Note that the metric is not conformal
to a Bianchi V metric.
It only admits a two--dimensional group of symmetries.
\section*{Appendix}
\setcounter{equation}{0}
\appendix
\def\theequation{\thesection{A}\arabic{equation}}
The real three--dimensional Lie algebra of Bianchi type $V$ is defined by
commutation relations
\[
[X_0, X_1]=X_1, \quad [X_0, X_2]=X_2, \quad [X_1, X_2]=0.
\]
We can choose its representation by 
$3 \times 3$ matrices 
\[ 
X_0= \left(
\begin{array}{ccc}
1  & 0  & 0  \\
0 & 0  & 0  \\
0 & 0  & 0  
\end{array}
\right), \quad 
X_1 = \left(
\begin{array}{ccc}
0  & 1  & 0  \\
0 & 0  & 0  \\
0 & 0  & 0  
\end{array}
\right), \quad
X_2 = \left(
\begin{array}{ccc}
0  & 0  & 1  \\
0 & 0  & 0  \\
0 & 0  & 0  
\end{array}
\right).
\]
The corresponding Lie group $G$ is the multiplicative group of real matrices of the form
\[ 
{\tt g} = 
\left( \begin{array}{ccc}
\rho  & x^1  & x^2  \\
0 & 1  & 0  \\
0 & 0  & 1
\end{array}
\right), \quad \mbox{where}\quad \rho\in\R^+, (x^1, x^2)\in \R^2.
\]
The left-invariant one-forms $\{ \ll^j, j=0, 1, 2 \}$ corresponding
to a basis $\{ X_j \}$ of a Lie algebra  are given by
\[ {\tt g}^{-1} d{\tt g} = \ll^j X_j, \quad \mbox{where} \quad {\tt g} \in { G}.\]
Hence 
\[
\ll^0 = \rho^{-1} d\rho, \quad  \ll^1 =\rho^{-1} dx^1,\quad  \ll^2 =\rho^{-1} dx^2,
\]
and
\be
\label{appendixa}
d\ll^0=0, \quad d\ll^1=\ll^1\wedge \ll^0, \quad d\ll^2=\ll^2\wedge\ll^0.
\ee
The left invariant vector fields defined by $L_j\hook\ll^k={\delta_j}^k$
are found to be
\be
\label{left_inv}
L_0=\rho\frac{\p}{\p \rho},\quad L_1=\rho\frac{\p}{\p x^1},\quad L_2=\rho\frac{\p}{\p x^2}.
\ee
The right--invariant one forms and vector fields can be found analogously from
$d{\tt g} {\tt g}^{-1}$. The algebra of isometries of metrics from Theorem \ref{complexV}
is spanned by the right--invariant vector fields, which in our coordinate basis are given by
\be
\label{right_inv}
R_0=\rho\frac{\p}{\p\rho}+x^1\frac{\p}{\p x^1}+
x^2\frac{\p}{\p x^2}, \quad R_1=\frac{\p}{\p x^1}, \quad  R_2=\frac{\p}{\p x^2}.
\ee
The left and right invariant vector fields satisfy the commutation relations
\begin{eqnarray*}
&&[L_0, L_1]=L_1, \quad [L_0, L_2]=L_2,\quad [L_1, L_2]=0,\\
&&[R_0, R_1]=-R_1, \quad [R_0, R_2]=-R_2,\quad [R_1, R_2]=0,\\
&&[R_j, L_k]=0.
\end{eqnarray*}

\end{document}